# Integrating Remote Sensing, GIS and Prediction Models to Monitor the Deforestation and Erosion in Peten Reserve, Guatemala

R. Bruno[1], M. Follador[1,2], M. Paegelow[2], F. Renno[2], N. Villa[3]

1 DICMA, Engineering, Università di Bologna, Italy
2 GEODE/CNRS, Université de Toulouse le Mirail, France
3 GRIMM, Université de Toulouse le Mirail, France
Corresponding author: marco.follador@mail.ing.unibo.it

ABSTRACT: This contribution provides a strategy for studying and modelling the deforestation and soil deterioration in the natural forest reserve of Peten, Guatemala, using a poor spatial database. A Multispectral Image Processing of Spot and TM Landsat data permits to understand the behaviour of the past land cover dynamics; a multi-temporal analysis of Normalized Difference Vegetation and Hydric Stress index, most informative RGB (according to statistical criteria) and Principal Components, points out the importance and the direction of environmental impacts. We gain from the Remote Sensing images new environmental criteria (distance from roads, oil pipe-line, DEM, etc.) which influence the spatial allocation of predicted land cover probabilities. We are comparing the results of different prospective approaches (Markov Chains, Multi Criteria Evaluation and Cellular Automata; Neural Networks) analysing the residues for improving the final model of future deforestation risk.

## 1. Introduction

The Peten is included into the Maya Biosphere Reserve, a part of the largest continuous tropical forest remaining in Central America. We studied the region of *La Joyanca* (archaeological site), in the North-West of Peten. After the collapse of Maya civilization the region was depopulated; from 1988 there was a new progressive immigration of Ladinos (driven by governmental plans and poverty of surrounding areas) with the first settlements in North side of Rio San Pedro. The subsequent fast deforestation, obtained by the traditional *Milpa* (slash and burn) is due to agriculture and pasture activities (Fig.1).

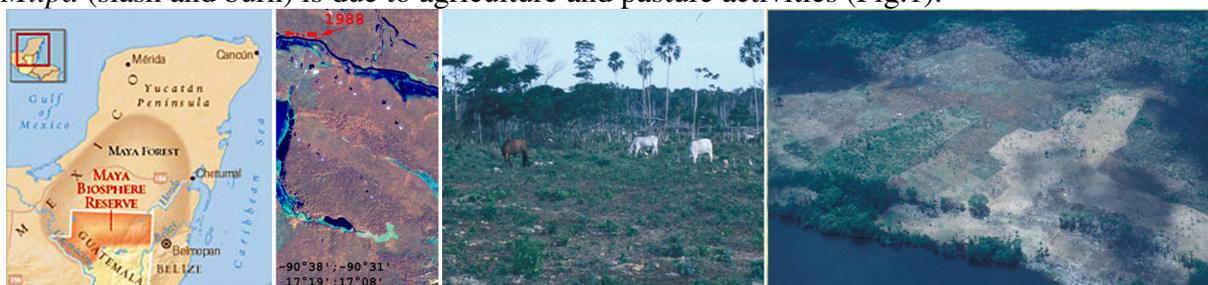
Fig. 1. La Joyanca region, NW of Petén. The effects of demographic raise became evident after 1988.

## 2. Methodology
### 2.1. Image Pre-processing

Four dates of Landsat TM imagery (1988, 92, 2000, 03; p20r48) and one Spot image (1998) were acquired. This Remote Sensing information was collected in different months and seasons, so it wasn't possible reduce the scene-to-scene variation due to atmospheric condition and phenology; the images of 1988 and 92 is representative of rain season, and the others of dry season. A relative radiometric correction based on DOC (dark object subtraction) were applied to reduce the atmospheric scattering within each scene (Chen *et al.* 2005); the water bodies *Laguna Tuspan* and *Agua Dulce* was chosen as reference. Binary mask was created to isolate water, clouds, cloud shadows and perennial wet lands (*Sibal*), considered as no important to deforestation analysis. The statistical study of pixel values (ND) permitted to calculate the more informative dataset reducing the between-band correlation and data volume; we used the Optimum Index Factor (based on ratio between the standard deviation and correlation index of image NDs) to chose the best Colour Composite for each year. We point out that the most informative RGB was composed by one visible, one near infrared and one mid infrared band (i.e. 1988→TM 145, 1992→TM 147); at the same time the less informative dataset was the so called "true colour composite" ( i.e. RGB→TM 321).

### 2.2. Spectral Indices and Change Detection Methods

We studied the ND point density maps combined different bands; the relationship between the red, near infrared, synthesized by NDVI (Normalized Difference Vegetation Index) was used to describe the forest clearing and regrowth; the relationship between the near infrared and mid infrared, synthesised by NDII (Normalized Difference Infrared Index), points out the eroded or hydric stressed areas (Fig.2).

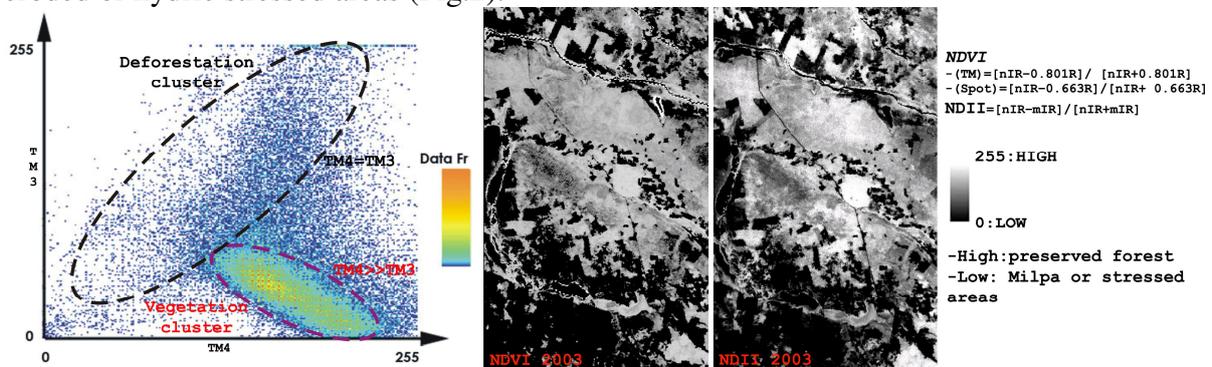

Fig. 2. NDVI and NDII used for deforestation analysis.

The NDII shows a higher resolution and permits a better definition of cleared and stressed areas; for the following temporal classification we used a new intermediate image (NDIm: Normalized Difference Index medium) built combined the NDVI and NDII. A change detection method based on RGB-(NDIm) was applied to time-series TM and Spot dataset. In particular we concentred our attention on the land use dynamics after the big wild fires in 1998; a RGB was built using the NDIm of 1998 (Spot)→R, 2000 (ETM)→G and 2003(ETM)→B. All images have an enhanced spatial resolution of pixel (5x5m). The interpretation of the final colour composite was assisted by the opinions of experts on this region. The values of NDIm images were standardized into a progressive scale (low-medium-high) to highlight the spatial distribution of vegetal vigour over the study area (Follador *et al.* 2006). The absence of temporal regularity in our data base does not allow eliminating the phenology and season effects; so we had some problems in separating the "yellow tones" due to deciduous plants cycle from the "yellow tone" due to atrophic impact. However this appraisal limit is only concentrated in a well defined region and it is quite simple to divide the two different dynamics using additional information (phyto geographic map, environmental criteria, etc.). Finally we recognized 10 categories, with different changes in vegetal vigour or different hydric stress, from 1998 (before big wildfires) to 2003; we show only the main dynamics, including classes with similar values and behaviour (i.e. white colour and clear rose colour have NDIm values minimally different, due to different humidity→ white colour) in the same group (Fig.3).

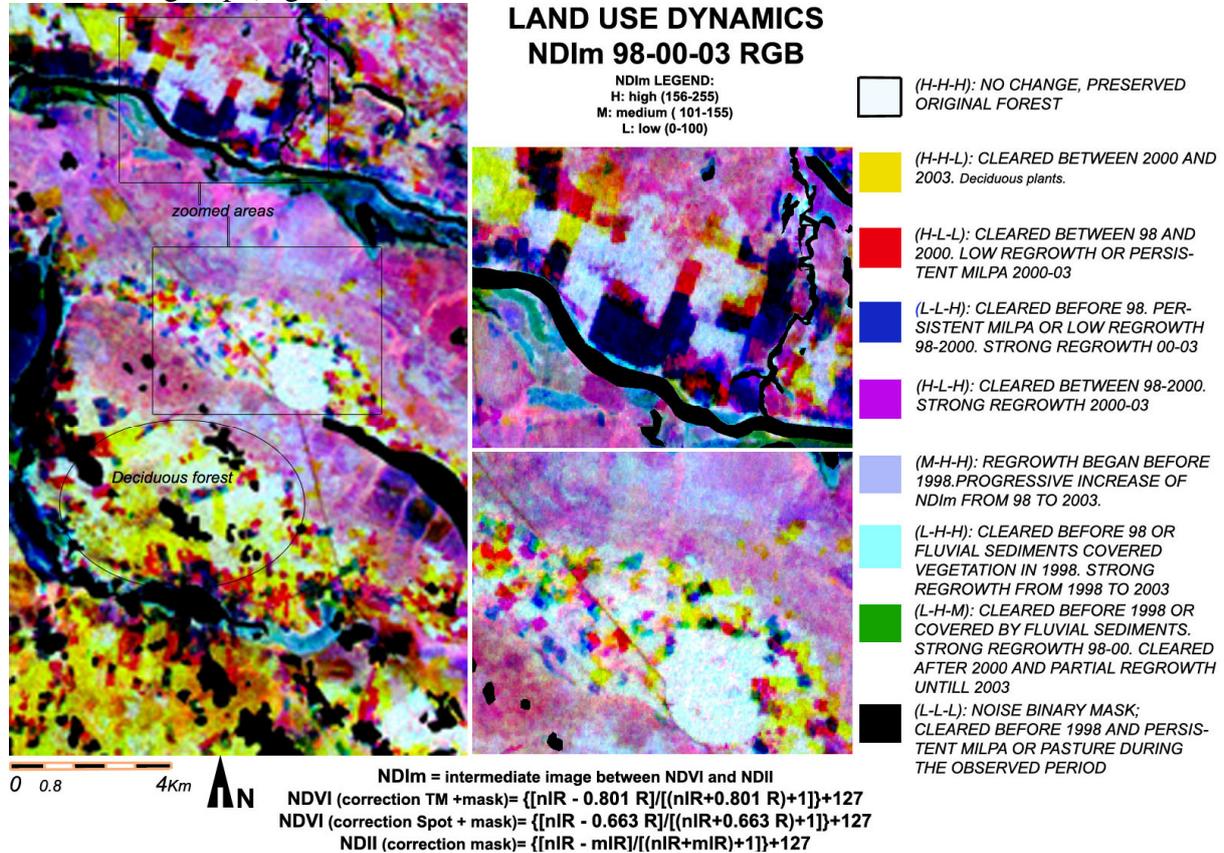

Fig. 3. Land Use dynamics of La Joyanca region, Peten, from 1998 to 2003. RGB-NDIm change detection method. Masked images (clouds, water, clouds shadows).

The results were checked by historical information of precedent field works (Métailié *et al.* 2000) and old aero-photos (visual interpretation method developed by Cohen *et al.* (1998), showing an high accuracy.

### 2.3. Images classification and extrapolation of environmental criteria

Now we prepare the information to introduce in our predictive models for simulating the development of deforestation for the next future. A supervised classification was applied to more informative Colour Composite (according to the OIF) for each date. We used a mixed algorithm based on MaxLikelihood (which analyzes only the ND of each single pixel) and on Interacted Conditional Model (for considering the data spatial distribution); so the class attributed to a pixel depends both from its value and from the classes of neighbouring pixels.

Initially we have pointed out 4 classes ( High Forest, Low Forest + Secondary Forest regrowth , Wet Lands, Milpa-Pasture + low regrowth) with a Kappa index  approximately of 0.7-0.8, due to the big confusion between High Forest and Low Forest, in particular during the rain season. We decided to reduce our study to 2 classes (Forest and Milpa-Pasture) and include the Wet Lands in the Binary Mask, focalizing the attention on farming and ranching impact of original ecosystem; the confusion matrix was calculated and the Kappa index pointed out an high accuracy (>0.9). From remote sensing images we extrapolated the main factors which influenced the deforestation phenomena: distance from San Pedro River (main way of access to forest), distance from the Pipeline (built in 1996), from roads (built in the last years) and DEM (derived from Shuttle Radar image). At the end we integrated the socio-economics information for "humanizing the pixel" and better explain the future land use dynamics. The Peten government has assigned a part of studied region to native tribes (vectorial mask of tribes areas borders) that deforested small areas for well adapted agriculture, while in the surrounding lands predominate the ranching activities (large cleared areas in progressive expansion); we also considered the different economic and agriculture potentiality of soils (correlated to geomorphology and pedology) because it is a good indicator of deforestation dynamics (agriculture➔small areas, pasture➔large areas) spatial distribution and subsequent soils stress.

**2.4. Neural Network models**
We used, for modelling this problem, one of the numerous neural tools developed during the past years: the multi-layer perceptrons. They were the first neural network models built in order to simulate the human brain but their applications are now far away from this original purpose. The use of multi-layer perceptrons is motivated by their great ability to approximate almost every function and, then, to adapt themselves to any problem where some classical statistical tools (linear models for examples) could fail. For more details about their mathematical properties and how they have been used in statistics, we advise to refer at Davalo and Naim (1969) or Bishop (1995).

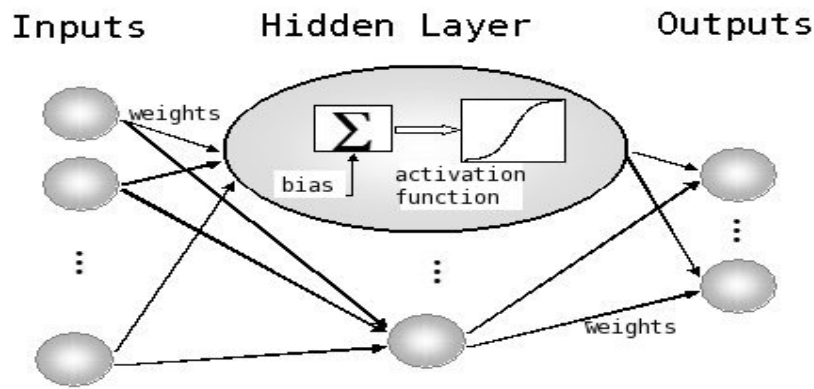

Fig. 4. One hidden layer perceptron.

We used a one hidden layer perceptron like shown in Fig.4. The predictive variables (land cover at previous dates, environmental criteria, etc.) are the "inputs" of the model (x). Each of them is used to calculate the hidden layer and the "outputs":

$$\psi_w(x) = \Sigma_{i=1}^{q} w_i^2 \, g\left(x^T w_i^1 + w_i^0\right)$$

where "q" is the number of neurons on the hidden layer, "w" are the weights and "g" is the activation function, usually the sigmoid:

$$g: z \rightarrow \frac{1}{1+e^{-x}}$$

"q" and "g" are chosen by experts whereas the weights are chosen automatically such that the outputs are as closed as possible to the target (the land cover at a fixed date) during a training step. Once the model has been trained with known examples (inputs/outputs: 1988/1992 and 1992/2000 maps), it is then used to predict the land cover at 2003 using the land cover map 2000 (new inputs). This predictive map can be compared with reality and the residuals analysed.

### 2.5. Geomatic model

Unlike pure mathematic models, geomatic prediction models applied to environmental dynamics include a part of human performed geographic analysis, to carry out the relationship between land cover dynamics and potential explanatory criteria. Among the multiple methodological approaches for predictive simulation in geomatics (Coquillard and Hill 1997) we used a combination of three modelling tools: a multi-criteria evaluation (MCE) to perform suitability maps for each category of the variable to be modelled, Markov chain analysis for prediction and, finally, in integrating step using MCE suitability scores for spatial implementation of Markovian conditional probabilities. This latest step arbitrates by multi-objective evaluation and cellular automata for realistic landscape pattern.

The prediction model is stochastic, handles with discrete time and finite states of land cover (modelled variable). To do so we use available GIS software components (implanted in Idrisi 32 Kilimanjaro) and a restrictive list of criteria so that the methodology would be easy to apply to other terrains. The calibration will be performed by modelling a known land cover state, the last available date. Therefore we use as training data the two earlier land cover layers and known and relevant environmental and social criteria. Validation will be obtained by comparison with a later, also known – but not used for predictive modelling – land cover state. The chosen approach may be considered as a "supervised" model with manual establishment of a knowledge base in comparison to "automatic" approaches like neural networks.

<u>Multi-criteria evaluation (MCE)</u>

The knowledge about former dynamics is essential to attempt the prediction of the future evolution or to build prospective scenarios (decision support). Therefore any model has to be supplied with values of initial conditions. In this study we consider two earlier land cover states as training dates to initialize the model. Performed values materialize statistically improved knowledge about land cover behaviour in space and time. The values are the two training land cover maps (depending variable) used to perform time transition probabilities and land cover relevant criteria (independent variable) correlated to land cover. The statistical tests (logistic regression, PCA) helped us to choose the criteria. The criteria might be split up into Boolean constraints and factors which express a land cover specific degree of suitability, variable in space. The constraints will simply mask space while the factors may be weighted and allowed to trade-off each other. Because each factor is expressed in proper units they have to be standardized to become comparable. Standardization signifies the recoding of original values (degrees, meters, per cent) to suitability values on a common scale reaching from 0 to 255 (best suitability). Based on statistical tests, recoding is processed by different ways: manual or by fuzzy functions. The factors, once standardized, are weighted by pairs using Saaty matrix (Saaty 1977) and performing the eigenvector. A second set of context-depending weights allows choice of risk and trade off levels.

<u>Markov chains – time transition probabilities</u>

To perform land cover extrapolation, we use Markov chain analysis (MCA), a discrete process with discrete time which values at instance $t_{+1}$ depend on values at instances $t_0$ and $t_{-1}$ (Markov order 2). The prediction is given as an estimation of transition probabilities.

MCA produces a transition matrix recording the probability that each land cover category will change to every other category and the number of pixels expected to change. The algorithm also generates conditional probability maps for each land cover showing the probability with which it would be found at each pixel after a specified number of time units.

<u>Integrating step based on multi-objective evaluation and cellular automata</u>

The spatial allocation of predicted land cover time transition probabilities uses MCE performed suitability maps and a multi-objective evaluation (MOE) arbitrating between the set of finite land cover states. Finally we add an element of spatial contiguity by applying a cellular automaton (contiguity filter). The algorithm is iterative so as to match with time distances between $t_{-1} - t_0$ and between $t_0 - t_{+1}$.

## 3. Conclusions

The study of past land use dynamics (1988-2003) in la Joyanca Region, Peten, shows two different trends: into the tribes' concession borders we can observe the birth of small cleared areas which periodically change, permitting a fast forest regrowth and a quite well adapted use of nature products; contrarily, in the surrounding regions, theoretically protected by Government laws, we point out a progressive expansion of deforestation for illegal ranching activities and subsequent stress and impoverishment of soils, with isolated phenomena of erosion. The predictive models permit to draw future scenarios of land use dynamics in La Joyanca Region; these represent an important tool for testing the effect of new Governmental policy of communitarian concessions to native tribes, which would have to guarantee a decrease of spontaneous and uncontrolled colonization and wild fires, often in the Natural Parks (Nittler *et al.* 2005). If the results of our predictive models will be encouraging, the policy of communitarian concessions would be extended to other site in the North of Peten, guaranteeing a better sustainable use of the tropical forest and replacing the obsolete and ineffective laws for Natural Reserves protection.